\documentclass{emulateapj}

\usepackage{threeparttable,textcomp,multirow}
\usepackage{float}
\usepackage{tensor}
\usepackage{graphicx}
\usepackage{color}
\newcommand{\simgt}{\,\hbox{\lower0.6ex\hbox{$\sim$}\llap{\raise0.6ex\hbox{$>$}}}\,}

\shorttitle{A theoretical color-velocity correlation for supernovae
  associated with gamma-ray bursts}
\shortauthors{Rapoport et al.}

\begin{document}

\title{A theoretical color-velocity correlation for supernovae
  associated with gamma-ray bursts}

\author{Sharon Rapoport\altaffilmark{1,2}, Stuart
  A. Sim\altaffilmark{1,2}, Keiichi Maeda\altaffilmark{3}, Masaomi Tanaka\altaffilmark{4} ,Markus Kromer\altaffilmark{5}, Brian
  P. Schmidt\altaffilmark{1,2}, Ken'ichi Nomoto\altaffilmark{3}}
\altaffiltext{1}{Research School of Astronomy and Astrophysics, The Australian  National University, Weston Creek, ACT 2611, Australia}
\altaffiltext{2}{ARC Centre of Excellence for All-sky Astrophysics (CAASTRO)}
\altaffiltext{3}{The Institute of Physics and Mathematics of the Universe, University of Tokyo, Japan}
\altaffiltext{4}{National Astronomical Observatory, Mitaka, Tokyo, Japan}
\altaffiltext{5}{Max Plank Institute for Astrophysics, Garching, Germany}

\begin{abstract}
We carry out the first multi-dimensional radiative transfer calculations to
simultaneously
compute synthetic spectra and light curves for models of supernovae driven 
by fast bipolar outflows.
These allow us to make self-consistent predictions for the orientation dependence of both colour evolution and spectral features.
We compare models with different degrees of asphericity and metallicity and
find significant observable consequences of both. In aspherical models,
we find spectral and light curve features that vary systematically
with observer orientation.
In particular, we find that the early phase light curves are brighter
and bluer when viewed close to the polar axis but that the peak flux is 
highest for equatorial (off-axis) inclinations.
Spectral line features also depend systematically on observer orientation,
including the velocity of the Si~{\sc ii} 6355\AA \ line. 
Consequently, our models predict a correlation between line velocity and
color that could assist the identification of supernovae associated with
off-axis jet-driven explosions.
The amplitude and range of this correlation depends on
the degree of asphericity, the metallicity and the epoch of observation 
but we find that it is always present and acts in the same direction.
\end{abstract}


\section{Introduction} \label{intro} 
Gamma-ray bursts (GRBs) provide a powerful tool to study galaxy
formation at high redshifts (e.g. \citealt{Conselice2005}) and provide a unique opportunity 
to observe extreme physical processes (e.g. \citealt{Bucciantini2009,WoosleyHeger2006}). 
Long duration GRBs (LGRB) have been shown to be connected with
broad-lined Type Ic supernovae
\citep[SNe~Ic][]{Galama1998,Iwamoto1998,Kulkarni1998,Bloom1999,Meszaros2006,Lindner2010}, 
implying that they are linked to the collapse of a massive star, see
\cite{Woosley2006} for a review. 
However, the exact nature of this connection is not yet understood, and is
especially poorly constrained in the very bright GRBs used to study
the high-redshift universe. 

The collapsar model \citep{Woosley1993,MacFadyen1999}
predicts that when the iron core of a massive rotating
star collapses into a black hole (BH), material accretes onto
the BH via a rotating disk. A narrow, highly relativistic 
 jet is launched at the black hole - accretion disk boundary which gives
rise to the observed gamma-ray burst. In addition, a sub-relativistic wind is 
driven off the disk, which drives the explosion of the star. Nucleosynthesis
in the explosion produces $^{56}$Ni, the decay of which powers the SN
emission \citep{Iwamoto1998}.
Angular momentum is believed to be
a crucial factor in determining the mass that collapses into the BH. A
more massive BH implies lower density in the accretion disk, which
might disfavour production of a GRB \citep{Woosley2007}. As higher metallicity
leads to greater mass loss via stellar winds, which causes the star
to lose angular momentum, GRB progenitors are expected to
have low metallicity \citep{Heger2003,Yoon2005}. In the close binary
progenitor scenario, the metallicity dependence is much weaker (e.g. \citealt{Fryer2007})

Observationally, this collapsar model implies that emission produced in 
the highly collimated jet is strongly affected by 
relativistic beaming such that it is only detectable when
the Earth is closely aligned with the jet direction. 
Therefore, detection of a GRB immediately suggests that the event is
observed pole-on (along the rotation axis).
In contrast, the associated SN could be
observed from all orientations, although its properties can be expected to 
depend on 
viewing-angle.

Studying the SN associated with LGRBs is one means to better
understand the explosion mechanism. Analysis of SN 1998bw, the first
SN believed to host a LGRB, supports the high-mass collapsar model
\citep{Iwamoto1998,Woosley1999,MacFadyen1999} and gives an estimated main sequence
progenitor mass of $\sim 30-40M_\odot$. However, this object's
$\gamma$-ray luminosities were orders of magnitude lower than the
typical GRBs studied at high redshifts and it is thus not typical of
the class. Subsequent modelling of a variety of SNe associated with GRBs
has revealed a wide distribution in the kinetic energy ($E_k$), ejecta
mass ($M_{ej}$) and $^{56}$Ni mass ($M_{56}$) parameter space (e.g. \citealt{Nomoto2004,Nomoto2006a}). For
example, in 2003 SN2003dh/GRB 030329 became the first bright GRB to 
have its SN studied in detail \citep{Hjorth2003,Stanek2003}. At redshift z=0.169
\citep{Greiner2003}, it was the first nearby 
GRB associated with a SN to have $\gamma$-ray and
afterglow properties similar to those observed for cosmological GRBs \citep{Price2003}.
Also, \cite{Berger2011} recently observed
the cosmological GRB/SN 2009nz, and reported broad line features with
lower velocities than 1998bw.
However, as explosion parameters are often estimated using the simple Arnett relation \citep{Arnett1982},
more detailed modelling is warranted (see discussion).

The collapsar scenario for LGRB-SN predicts that in many
cases we will be able to observe the SN explosion but will not see the
GRB because of our orientation. One way to test the model is to 
hunt for the predicted population of asymmetric SN explosions in which
our line-of-sight is off the jet axis. A promising approach is radio
surveys looking for emission associated with the interaction of
relativistic outflow with
a previously emitted wind or the circumstellar material. Based on a sample of 68 Ib/c SNe,
\cite{Soderberg2006} argued that $\lesssim $ 10\% of type Ib/c SNe are
associated with off-axis jets. However, radio observations have also
shown that outflows powered by a central engine may be present in some
objects for which no GRB was detected \citep{Soderberg2010}.

Other phenomena that could be explained as an off-axis GRB
include X-ray flashes \citep{Yamazaki2004}, which were also found to
sometimes be associated with SNe (e.g. XRF100316D/SN2010bh;
\citealt{Olivares2012,Bufano2011,Cano2011}, XRF060218/SN2006aj; \citealt{Campana2006,Pian2006}). 
Some evidence for jet-like SN explosions has also been uncovered via
late-time optical observations of stripped-envelope core collapse
SNe. Once the ejecta is optically thin the observed
blue- and red-shifted lines follow the shape of the
ejecta. In this stage, double-peaked line profiles arise when 
a bipolar SN is observed
off-axis \citep{Maeda2002},
and some cases were detected and analysed
by, for example, \cite{Maeda2008} and \cite{Taubenberger2009}.

The aforementioned studies of radio emission and late-time optical spectra are powerful tools to search for evidence of jets having been 
involved in SN explosions. However, signatures are also expected in ultraviolet/optical/near-infrared emission around maximum light
of the SN explosion. Observations of this phase are typically more complex to interpret than e.g. late-phase nebular spectra but 
are extremely important since they are the most routinely accessible data. To interpret such data
requires suitable explosion models and synthetic spectra and light curves. 
The quest to understand the GRB-SN connection has led to the development of 
numerous hydrodynamical studies of
jet-driven models.\ \cite{Proga2003} and later \cite{Sawai2005} showed
via simulations that magnetohydrodynamical processes alone can launch and sustain polar outflows
from the accretion disk around a black hole formed by the collapse of the core of a massive star.
The success of such simulations of jet launching have ignited a study of theoretical modelling of
asymmetric SNe. In particular, hydrodynamical simulations of asymmetric explosion models have been conducted
\citep{Maeda2002,Maeda2006,Maeda2006b,Tanaka2007}, and compared with observations, 
mostly looking for a promising model to
explain the extraordinary SN 1998bw.

In this paper our goal is to carry out self-consistent, multi-dimensional, time-dependent radiative transfer simulations to predict
the angle-dependent spectra and
light curves for asymmetric Ic SNe. 
In previous studies \citep[e.g. ][]{Tanaka2007}, light curves were calculated with a very simplified treatment of opacity while spectra were computed separately using a time-independent approach.
Our new self-consistent calculations will
help quantify the observable signatures of a jet-driven explosion and, in particular, allow us to 
predict the differences in optical colors and spectral features that are expected for a jet-driven explosion viewed off-axis.
As input, we work with the asymmetric explosion models of
\cite[][, hereafter M02]{Maeda2002}. We also investigate the influence of
progenitor metallicity on the optical observables since several studies have suggested a possible link between the metallicity
of the SN environment and whether it will produce a GRB
\citep{Sollerman2005,Modjaz2008,Modjaz2006,Levesque2010}.
We begin (\S \ref{method}) by 
describing the explosion models and principles of our radiative
transfer code. In \S \ref{results} we outline our results and in \S
\ref{conc} we discuss our conclusions.

\section{Methods}\label{method}
\subsection{Explosion Models} 

For this study, we adopt the jet-driven explosion models of M02.
These were created by exploding the core of a star that had initial
mass of $40M_{\odot}$ \citep{Iwamoto1998}. A cut 
in mass coordinate (at 2.4~$M_{\odot}$) was used to divide the core  
model into an inner region, which is
assumed to collapse to a black hole, and an outer region that will form
the SN ejecta.
2D hydrodynamical and nucleosynthesis calculations were performed to
simulate the explosion of the outer layers. The explosions were driven
by inserting an excess of kinetic energy in the region below the mass
cut. This gives rise to a shock which propagates through the model,
unbinding the material and triggering nuclear burning. 
All models have a total ejecta mass of $\sim 10$~$M_{\odot}$, although small
deviations arise owing to the treatment of boundary conditions in the
hydrodynamic simulations. Different asphericities were induced in the models 
by varying the injection of the kinetic
energy. Specifically, direction-dependent initial velocities were imposed
such that the velocity along the polar (i.e. assumed jet) axis ($v_z=\alpha z$) 
was larger than that along the equatorial direction ($v_r = \beta r$). The degree of asphericity is conveniently parametrized 
by the ratio $\alpha / \beta$. 

The M02 models explored a range of values for the final kinetic energy and degree of 
asphericity. 
Since this study focuses on the effect of asphericity on the light curves and spectra, we will consider only a subset of their 
models that have a fixed final kinetic energy
of $E_{51}=20$
($E_{51}\equiv E_{iso}/10^{51}$ergs): models A, C and
F of M02. Of these, model A is the most aspherical, C
has intermediate sphericity and model F is spherically 
symmetric. Important parameters of these models, and their $^{56}$Ni yields,
are given in Table~\ref{tab1}.

Observations of SNe associated with LGRBs do not reveal any H or
He. Modelling of SN1998bw was successful when using as a progenitor a core stripped down to its O
layer by the time of explosion \citep{Maeda2002}. Therefore,
for our radiative transfer calculations the composition of all
unburnt material has been set to that of the O layer with a small
fraction ($\sim$1\%) of C (i.e. we do not include any He-rich material in the ejecta). 
To explore the effect of progenitor metallicity, we have included
all abundances of elements heavier than Na (up to atomic number 30)
in the composition of the unburnt material. We adopted the solar
elemental abundances of \cite{Anders1989} and have computed synthetic observables for models 
with metallicities (relative to solar) of $Z = 0.1Z_\odot$ and
$Z=1Z_\odot$. Note that the solar element abundances have only been added to
the input conditions for our radiative transfer simulations -- i.e. we explore
the effect of metallicity in the unburnt material on the synthetic
spectra and light curves but do not investigate how progenitor
metallicity influences either the hydrodynamics or nucleosynthesis of
the burnt material. In addition, the metallicity of the
primordial material of a star will influence its evolution (e.g. \citealt{Meynet2009}). This will effect the abundances and stellar
structure at the time of collapse, which will alter the expected SN
spectra and light curve. We plan to investigate the expected
variations in a coming paper (Rapoport et al. in prep) in which we consider a wider range of progenitor models.  
Throughout this paper, we will denote our calculations adopting solar
metallicity for the M02 models A (C, F) as A1 (C1, F1) and those for
sub-solar metallicity ($Z=0.1Z_\odot$) as A0.1 (C0.1, F0.1).

\begin{figure}[htbp!]
\includegraphics[width=\columnwidth]{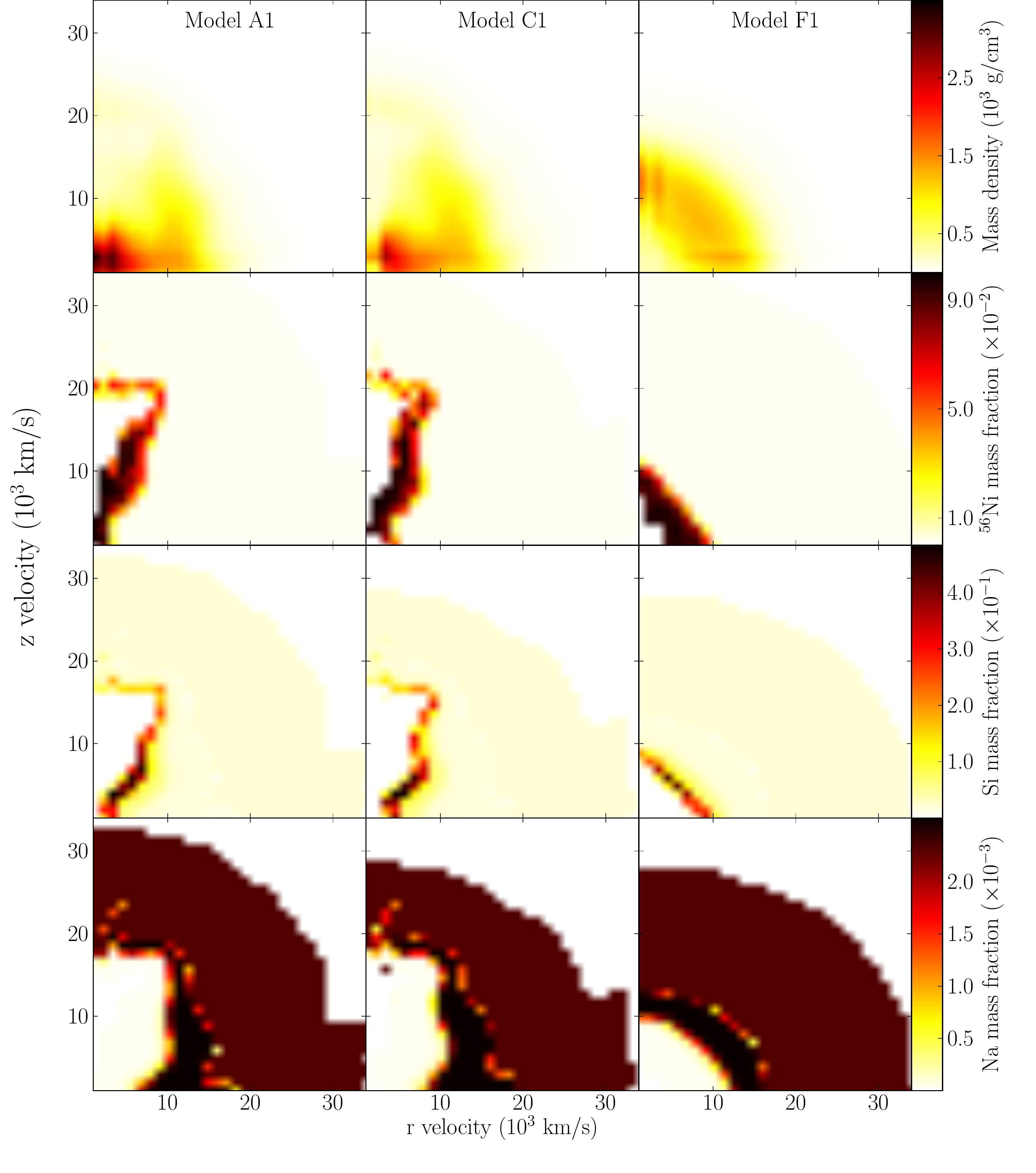}
\caption[short]{Left column: Model A1, middle column: Model C1, right column: Model F1. From top
  to bottom: Total mass density, $^{56}$Ni, Si and Na mass
  fractions. While model F is spherically symmetric the 'line' shape
  seen for the Si and $^{56}$Ni mass fractions is due to numeric
  effect as hydrodynamic simulations tend to give artefacts near the
  imposed axes of symmetry. The bulging along the equatorial is
  another simulation feature due to numeric effects and the colliding at the equator of the
  bow shocks which are driven by the jet.}
\label{fig:model}
\end{figure}

Fig. \ref{fig:model} illustrates the spatial distribution of mass
density and the abundances for key elements that will feature in our
discussion. In the aspherical models (A1, C1), the explosion dynamics
gives rise to a relatively dense core region that is extended in the
equatorial direction. In contrast, the polar region (i.e. the
direction in which the high velocities were initially injected) has
relatively low density. $^{56}$Ni is produced by explosive Si burning and, 
in all aspherical models, its mass fraction is highest in the low density
regions. This gives a near-conical shape to the $^{56}$Ni
distribution for the most aspherical model (see also Fig. 2 in
\citealt{Maeda2006b}). When the degree of asphericity is reduced (model C1), 
the $^{56}$Ni distribution no longer flares out from the pole and gradually 
looks more like the spherically symmetric model (F1).

Si (and other intermediate-mass elements) is produced by explosive O
burning (e.g. \citealt{Nomoto2006a}). Its abundance distribution is similar in shape to 
$^{56}$Ni, being concentrated around the outer edge of the
$^{56}$Ni-rich region. Si is also present in small quantities in the
outer layers since we have assumed non-zero progenitor metallicities
in all models. The Na in our models is dominated by the contribution 
in the unburnt
material. Its abundance distribution is therefore comparatively uniform and
its concentration highest in the outer layers.

\begin{deluxetable}{crrcc}
\tabletypesize{\footnotesize}
\tablewidth{0pc} 
\tablecaption{Explosion Models\label{tab1}}
 \tablehead{
 \colhead{Model} &  \colhead{$M_{ej} (M_{\odot})$} &
 \colhead{$M_{56} (M_{\odot})$}&\colhead{$E_k (E_{51})$} & \colhead{$\alpha / \beta$\tablenotemark{1}}}
 \startdata
 A1 (A0.1) & 10.10 &0.23      & 20   & 16 \\
 C1 (C0.1) & 10.05 & 0.28      & 20   & 8 \\
 F1 (F0.1) & 10.08 & 0.38     & 20   & 1 
\enddata
 \tablenotetext{1}{Asphericity parameter --- see text for details}
\end{deluxetable}

\vspace{1cm}

\subsection{Radiative Transfer}
Our radiative transfer simulations were carried out using \emph{ARTIS}
(\citealt{Sim2007,Kromer2009}), which is a Monte Carlo (MC), multi-dimensional, time-dependent, 
radiative transfer code 
based on the methods described by
\cite{Lucy2002,Lucy2003,Lucy2005}. The code computes synthetic spectra for SN
explosion models 
assuming homologous expansion of the ejecta and the Sobolev
approximation for line opacity. 
MC radiative transfer methods 
has been applied to a variety of SN
explosion models and tested by comparison to other calculations in e.g. \cite{Kasen2006} (using the \emph{SEDONA} code) and \cite{Kromer2009} (for {\emph{ARTIS}).
Here, for the first time, we apply \emph{ARTIS} to models for SNe~Ic.
We note that, for SNe~Ic explosions with a compact progenitor, homologous expansion 
is expected to be reached within the first $\sim$10 seconds
and is therefore a 
good approximation at the times of $\sim$days that will be studied here.).

As input, the code requires specification of the velocity, density and composition
for each grid cell (a Cartesian grid is used). The total
energy emitted from the decay chain
$^{56}$Ni$\rightarrow^{56}$Co$\rightarrow^{56}$Fe is determined from
the model and is divided into a set of MC quanta. At the start of the radiative transfer simulation, these quanta are placed in the model according to 
the $^{56}$Ni densities in the form of radioactive pellets
\citep{Lucy2005}.
During the simulation these convert into photon
packets in accordance with the radioactive decay times. The photon
packets then propagate through the model of the ejecta and can interact with the
medium via Compton scattering, pair production, bound-free and
bound-bound absorption. For bound-bound transitions we use a line
  list extracted from the data of \cite{Kurucz1995} CD23, as describe in
  \cite{Kromer2009}.
The time, direction and frequency of escaping
energy-quanta are recorded to allow
angle-dependent spectra and light curves to be created. We refer the
reader to \cite{Kromer2009} for full details on the MC techniques
employed by \emph{ARTIS}. 

Spectra were generated for models A1, A0.1, C1, C0.1, F1 and F0.1. The
simulations were started at three days after explosion and were run until 50
days after explosion with 60 time steps, logarithmically spaced. While
a study of the models at earlier start time is appealing, the high
opacities immediately after explosion results in a heavy computation
requirement. 
In addition, this paper focuses on SNe associated with real GRBs,
where early time SN properties are practically undetectable under the
luminous GRB afterglow -- consequently predictions for very early times
are not easily testable.
A grid of 68$\times$68$\times$68 cells was used, giving a resolution of 
$\sim1,180$ km/s. As the explosion models are two dimensional with
symmetry along the jet direction, the synthetic observables depend only on the 
angle ($\theta$) between the observer line-of-sight and the polar axis
of the model. For the aspherical models, we extracted synthetic
observables for 5 values of $\theta$ by dividing the emergent MC
quanta into 5 bins of equal solid angle across the range $0 < \theta <
90$~deg. 

\section{Results}\label{results}

\subsection{Light curves}\label{light_curve_section}

The main advantage of our simulations compared to previous studies of
the M02 explosion models (\citealt{Tanaka2007}, hereafter T07) is a
self-consistent, time-dependent treatment of the spectral
evolution. This means we can extract light curves that are fully
consistent with both the spatial distribution of $^{56}$Ni in the
model and the detailed opacities needed to compute realistic spectra.  
Fig.\ \ref{fig:A_lc} shows light curves for the standard Bessell
\citep{Bessell2012} $U$, $B$, $V$, $R$ and $I$ filters\footnote{For
  convenience, the filter functions are shown in the upper panel of
  Fig.\ \ref{fig:spectra_features}.} for Model A1 for four different
observer orientations. We also show the bolometric $UVOIR$ light
curves, defined by a top-hat filter extending from 3,000\AA \ to 9,000\AA.

\begin{deluxetable}{rrrrr}
\tabletypesize{\footnotesize}
\tablewidth{0pc} 
\tablecolumns{5} 
\tablecaption{$V$ - band maximum light times in days after explosion \label{vband}}
\tablehead{ 
\colhead{}    &  \multicolumn{4}{c}{$V$} \\ 
\cline{2-5} \\
\colhead{Model}    & \colhead{$25^\circ$\tablenotemark{1,2}}  &
\colhead{$45^\circ$}    & \colhead{$60^\circ$}&\colhead{$85^\circ$}}
\startdata
 A1   & 22 & 24 &24&24\\
 A0.1 & 20&22&22&22\\
 C1   & 24&24&24&24\\
 C0.1 & 23&24&24&24\\
 F1   & \multicolumn{4}{c}{29} \\
 F0.1 & \multicolumn{4}{c}{28}\\

\enddata
\tablenotetext{1}{Values are quoted for four observer inclination angles (measured relative to the polar axis).}
\tablenotetext{2}{One number for all columns implies no angle
  dependency.}
\end{deluxetable}

In $B$, $V$, $R$ and $I$, the light curves for all observer
orientations reach peak around 20 -- 25 days after explosion, 
($\sim$7 days later than typical
for SNe associated with GRBs, see discussion) and then
decline monotonically. In these bands, the time of peak depends only
weakly on the observer orientation but tends to be earlier for
inclinations close to the pole (small $\theta$), particularly in the
bluer bands.  In $U$ band, the influence of orientation is much
stronger -- the $U$ light curve rises much more quickly and peaks
around ten days earlier for a polar compared to equatorial
inclination. The widely-used 
light curve decline rate parameter $\Delta m_{15}$ (the increase in 
magnitude from peak to 15 days after peak), is similar for the
asymmetric models and vary slightly with angles (see Table~\ref{tab:dm15}).  

\begin{deluxetable*}{rrrrrrrrrrrrrrrrrrrrrrrrr} 
\tabletypesize{\footnotesize}
\tablecolumns{25} 
\tablewidth{0pc} 
\tablecaption{$\Delta m_{15}$ - the increase in 
magnitude from peak to 15 days after peak in $U$, $B$, $V$, $R$ and $I$ bands for different observer inclination angles\label{tab:dm15}}
\tablehead{ 
\colhead{}    &  \multicolumn{4}{c}{$U$}&\colhead{}   & 
                      \multicolumn{4}{c}{$B$}&\colhead{}   & 
                      \multicolumn{4}{c}{$V$}&\colhead{}   & 
                      \multicolumn{4}{c}{$R$}&\colhead{}   & 
                      \multicolumn{4}{c}{$I$} \\ 
\cline{2-5} \cline{7-10} \cline{12-15} \cline{17-20} \cline{22-25}\\ 
\colhead{Model}    & \colhead{$25^\circ$\tablenotemark{1}}  & \colhead{$45^\circ$}    & \colhead{$60^\circ$}&\colhead{$85^\circ$} &
\colhead{}    &\colhead{$25^\circ$}   & \colhead{$45^\circ$}    &\colhead{$60^\circ$} &\colhead{$85^\circ$} &
\colhead{}    &\colhead{$25^\circ$}   & \colhead{$45^\circ$}    &\colhead{$60^\circ$} &\colhead{$85^\circ$} &
\colhead{}    &\colhead{$25^\circ$}   & \colhead{$45^\circ$}    &\colhead{$60^\circ$} &\colhead{$85^\circ$}&
\colhead{}    &\colhead{$25^\circ$}   & \colhead{$45^\circ$}    & \colhead{$60^\circ$}&\colhead{$85^\circ$}}
 \startdata
 A1   & 0.4 &1.0 &1.0&0.8&&\multicolumn{4}{c}{0.7\tablenotemark{2}}&&\multicolumn{4}{c}{0.9}&&\multicolumn{4}{c}{0.9}&&0.7&0.6&0.7&0.6 \\
 A0.1 & 0.6&1.0&1.1&1.0&&\multicolumn{4}{c}{1.1}&& \multicolumn{4}{c}{1.0}&&\multicolumn{4}{c}{1.0}&&\multicolumn{4}{c}{0.6}\\
 C1   & 0.9&0.7&0.9&0.9& &\multicolumn{4}{c}{0.7}&&0.9&0.8&0.8&0.8&&0.9&0.9&0.9&0.7&&0.7&0.6&0.6&0.6\\
 C0.1 & \multicolumn{4}{c}{1.0}&&1.1&1.0&1.1&1.0&&1.0&0.9&0.8&0.9&&0.9&0.9&0.9&0.7&&0.6&0.6&0.6&0.5 \\
 F1   & \multicolumn{4}{c}{1.0} &&\multicolumn{4}{c}{0.6}&&\multicolumn{4}{c}{0.6}&&\multicolumn{4}{c}{0.6}&&\multicolumn{4}{c}{0.4}  \\
 F0.1 & \multicolumn{4}{c}{1.2}&&\multicolumn{4}{c}{1.0}&&\multicolumn{4}{c}{0.7}&&\multicolumn{4}{c}{0.6}&&\multicolumn{4}{c}{0.4} 

\enddata
\tablenotetext{1}{Values are quoted for four observer inclination angles (measured relative to the polar axis).}
\tablenotetext{2}{One number for all columns implies no angle
  dependency.}
\end{deluxetable*}

In general, the pre-maximum light curves are brighter for polar
inclinations but around peak the equatorial light curves become
brighter. Since the angle dependence is strongest in the bluest bands,
there are also significant differences in the color evolution for
polar and equatorial lines of sight. This is illustrated in Fig.~\ref{fig:BMV_A}, which shows the $B-V$ color evolution for
Model~A1. If viewed close to the polar direction, the color becomes
dramatically redder during the evolution towards maximum light. In
contrast, the equatorial color evolution is much weaker, showing increasing
blue-ward evolution starting a few days prior to maximum light.
Similar orientation-dependent trends are predicted in our other
asymmetric models. The scale is smaller in the more spherical models
(in Model~C1 the difference between $B-V$ for polar and equatorial
viewing angle at maximum light is $~$0.1~mag.) and larger in 
corresponding models with
reduced metallicity
(e.g. Model~A0.1, where the difference is $~$0.45~mag.; 
see Section~\ref{met}). 

The origin of the angle-dependence in our light curves and colors can
be largely traced to the asymmetric distribution of iron group nuclei,
particularly $^{56}$Ni and its decay products
(see Fig.\ \ref{fig:model}), and the evolution
of the ionization state of the ejecta, which is shown in Fig.\
\ref{fig:ions}. At early times, the effective photosphere is located
at relatively high velocities. Since the $^{56}$Ni distribution is
most extended around the poles, it is there that the photosphere is
first most directly heated, leading to
initially stronger polar emission. The ionisation state is generally
high at early times (e.g. in the $^{56}$Ni-rich regions, Co is
dominated by Co~{\sc iii} as seen in Fig.~\ref{fig:ions}
\footnote{In Fig.~\ref{fig:ions}, we chose to show Co as an indicator
  for the typical ionization conditions for iron group elements
  synthezised in the explosion. For the epochs shown in the plot,
  $^{56}$Co is the most abundant iron-group isotope although we note
  that other iron group elements (including Ti, Cr and Ni) are still
  very important for line blocking in the blue.} 
),
such that this emission is fairly blue.
As the ejecta expands, the opacity of the outer ejecta drops and the
photosphere recedes deeper into the ejecta. Once radiation starts to
escape from the inner ejecta, the larger projected area of the
$^{56}$Ni-rich region for equatorial inclinations generally favours
brighter peak magnitudes for these orientations. 
At the same time,
the line blanketing in the blue is stronger along the poles where
metals synthesised in the explosion are present in the region around the
photosphere. This leads to relatively red colors.
This iron-group material is mostly singly- and doubly- ionised (upper
panels of Fig.~\ref{fig:ions}) but it gradually recombines, causing
the polar line blanketing to become even stronger with time in the
rise to maximum light. 

\begin{figure}
\begin{center}
\includegraphics[width=\columnwidth]{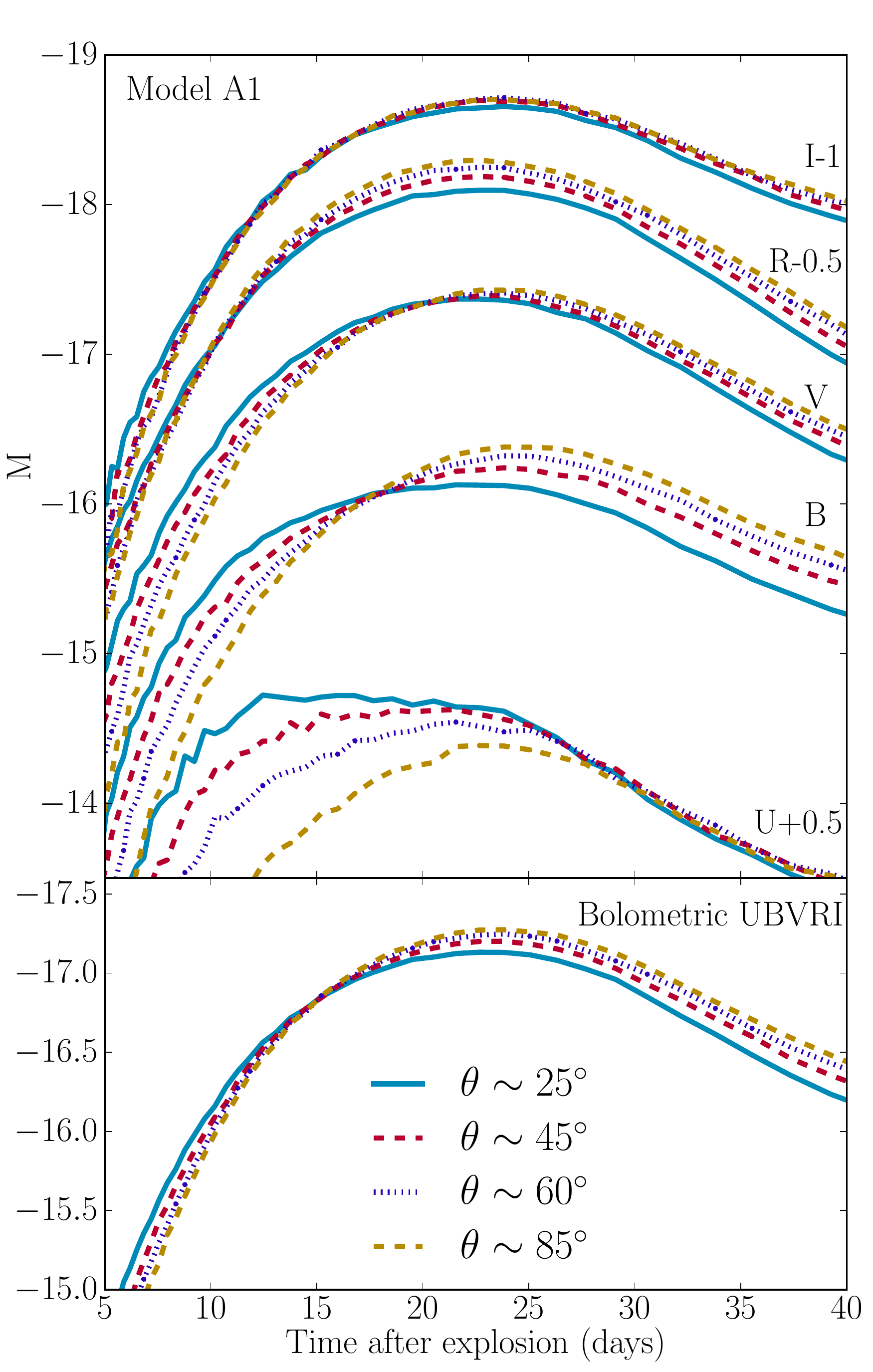}
\caption[short]{Light curves for model A1 for the different viewing
  angles. $\theta=0^\circ$ is the jet
direction and $\theta=90^\circ$ is the equator. See top panel of
Fig. \ref{fig:spectra_features} for the filter curves.}
\label{fig:A_lc}
\end{center}
\end{figure}

\begin{figure}
\begin{center}
\includegraphics[width=\columnwidth]{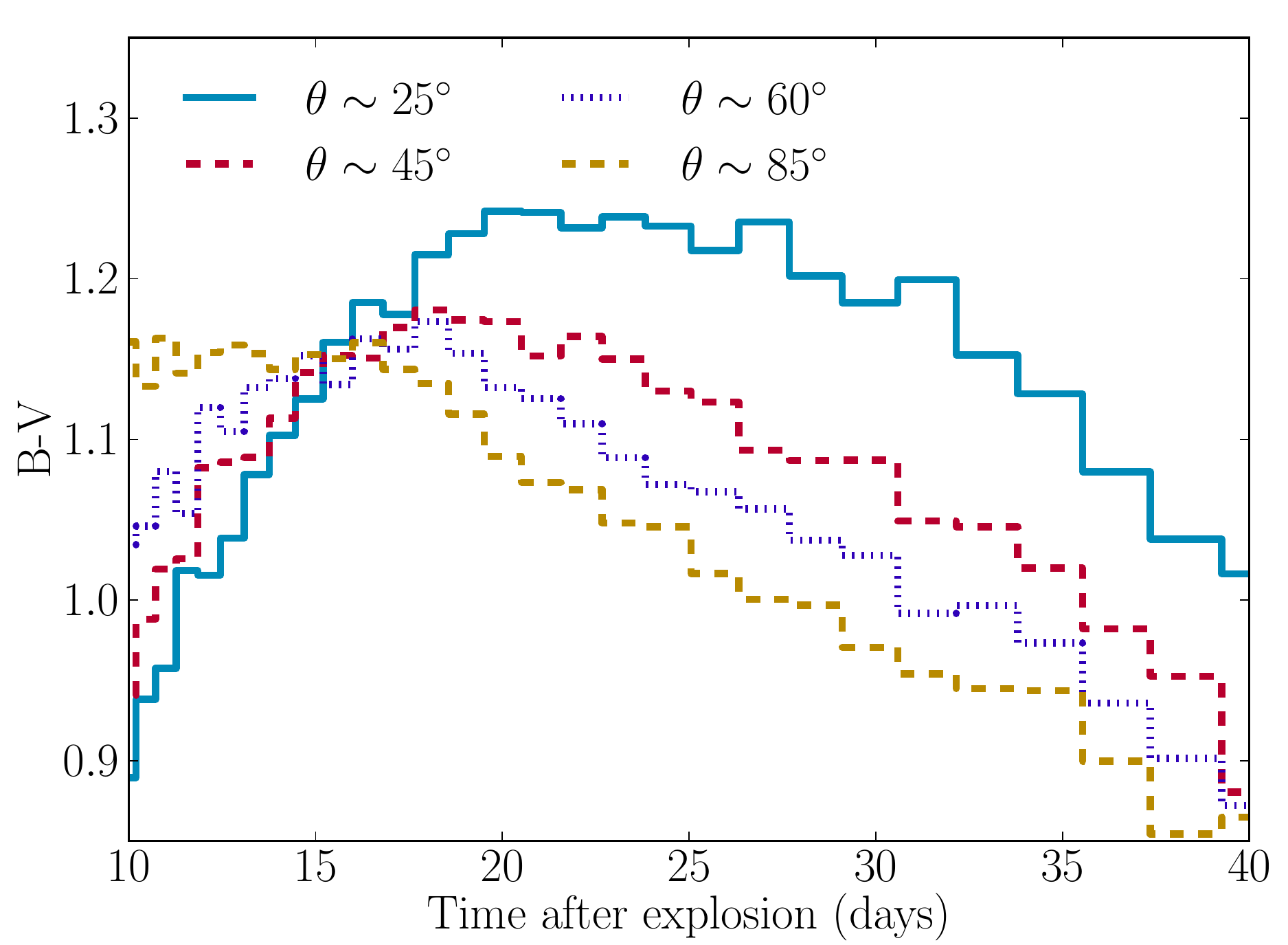}
\caption[short]{B-V color evolution for model A1 for the different viewing
  angles. $\theta=0^\circ$ is the jet
direction and $\theta=90^\circ$ is the equator. See top panel of
Fig. \ref{fig:spectra_features} for the filter curves.}
\label{fig:BMV_A}
\end{center}
\end{figure}

\begin{figure}
\begin{center}
\includegraphics[width=\columnwidth]{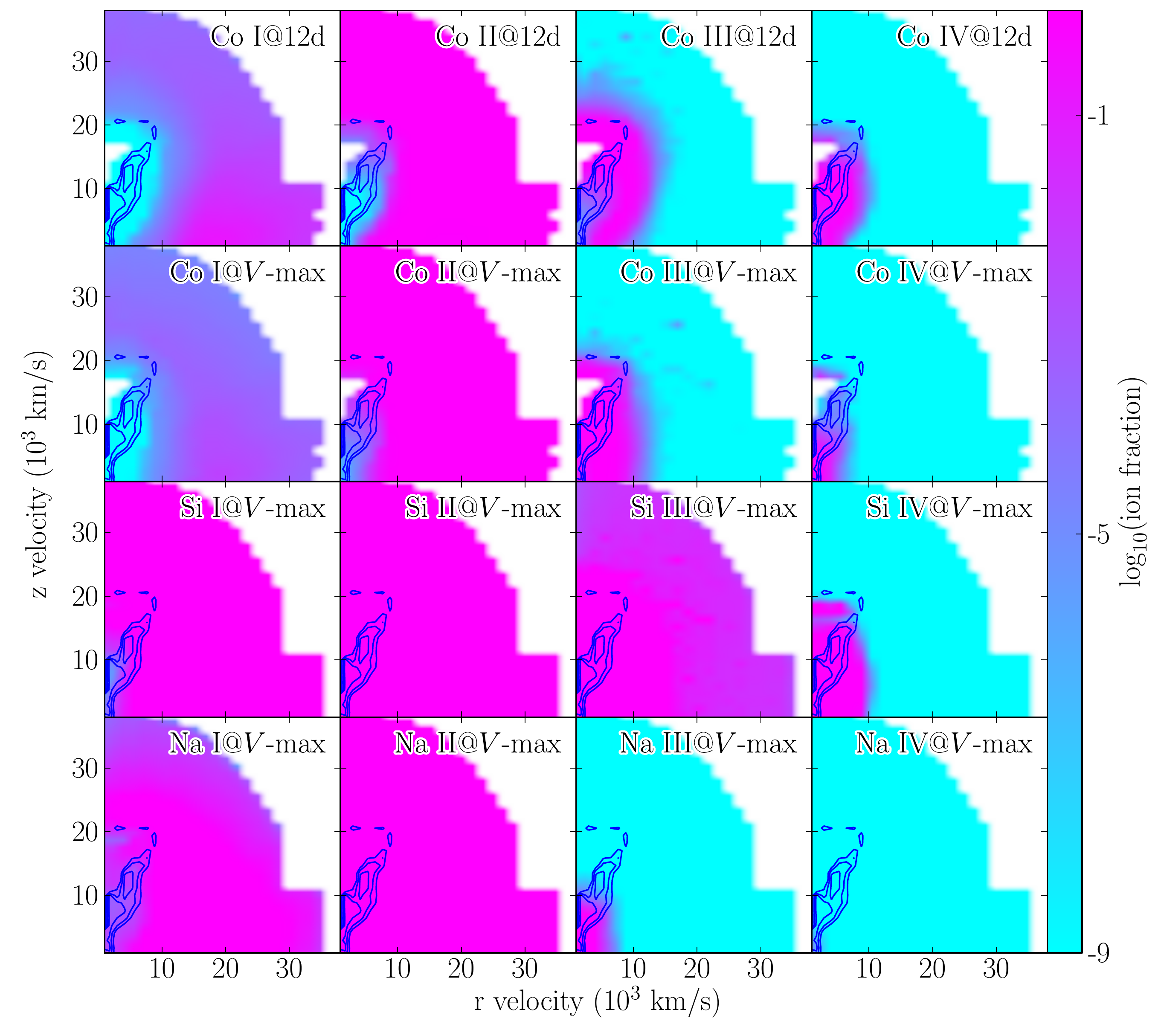}
\caption[short]{Ion fractions for model A1. From left to right the ions
  increase from {\sc i} to {\sc iv}, from top to bottom the ions are Co at 12
  days ($\sim$10 days before $V$-Band maximum lights), and Co, Si and
  Na at $V$-Band maximum light (see table \ref{vband} for details). The blue curves are contours of the
  $^{56}$Ni mass fraction and correspond to 0.05, 0.07 and 0.09.}
\label{fig:ions}
\end{center}
\end{figure}

\begin{figure}
\begin{center}
\includegraphics[width=\columnwidth]{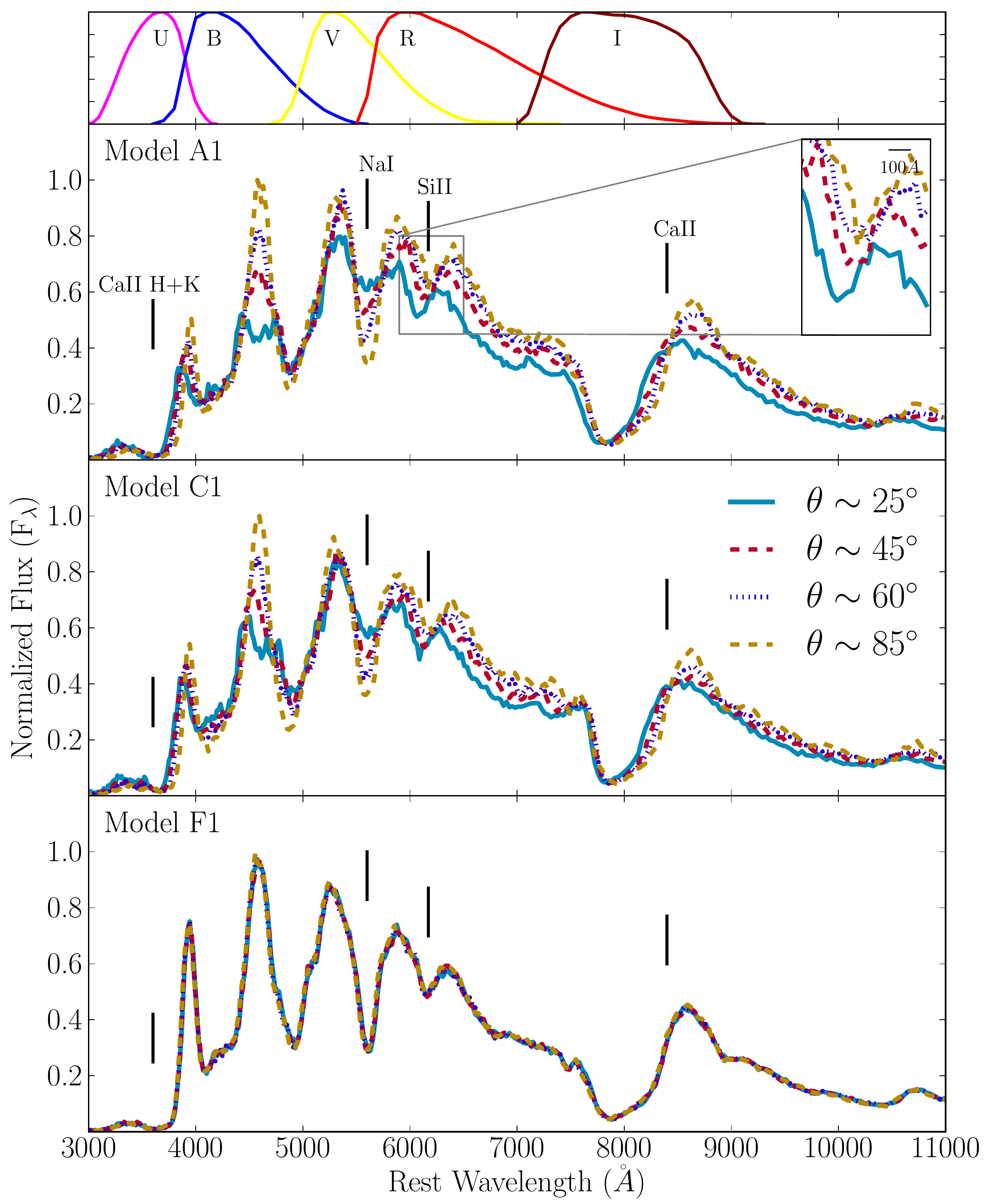}
\caption[short]{Synthetic spectra at $V$-band maximum (see table
  \ref{vband} for details) light for models
A1 (most asymmetric), C1 (intermediate symmetry) and F1 (spherically
symmetric, upper, middle and lower panels, respectively) for different
viewing angles. The top panel shows the filter functions used to generate the
light curves in Fig. \ref{fig:A_lc}. $\theta=0^\circ$ is the jet
direction and $\theta=90^\circ$ is the equator.}
\label{fig:spectra_features}
\end{center}
\end{figure}

\subsection{Si~{\sc ii} velocity and correlation with color}

The Si~{\sc ii} 6355\AA  \ transition is
an unmistakable absorption feature that is commonly seen in the red
part of SN spectra. 
The asphericity of the Si-rich region (see  
Fig. \ref{fig:model}) gives rise to a significant
angle-dependence of the Si~{\sc ii} 6355\AA \
absorption line velocity. 
This was previously reported by T07 and is confirmed in our study 
(see Fig.\ \ref{fig:spectra_features}, which shows the optical spectra computed
for our models at $V$-band maximum light).
To illustrate the origin of the angle-dependence of the Si line, 
Fig.~\ref{fig:SiII} shows the Sobolev optical depth of the 6355\AA  \
line at 15, 25 and 35 days for models A1, C1 and
F1 and indicates the location of the region of last interaction for
escaping $R$-band photons.
At all epochs, the region in which the line is optically thick extends
to significantly higher velocity along the polar than equatorial
direction (in models A1 and C1). Therefore the absorption extends to
higher velocity in the spectrum when observed pole-on. Notice,
however, that in most cases the region of last interaction for
escaping $R$-band photons does extend outside the region in which the
Si line is optically thick. Consequently, the Si line trough has
usually been partially refilled such that the line core does not
appear to be saturated. 

The same trend appears across the spectrum and can clearly be seen
in the Ca~{\sc ii} triplet, especially on the blue side of the emission
part. Higher velocities towards the equator are also seen in the
blue (e.g.$\sim$4400\AA), although the overlapping of many Fe, Ti and Co lines makes
quantitative analysis more complicated in this region.
 
\begin{figure}
\begin{center}
\includegraphics[width=\columnwidth]{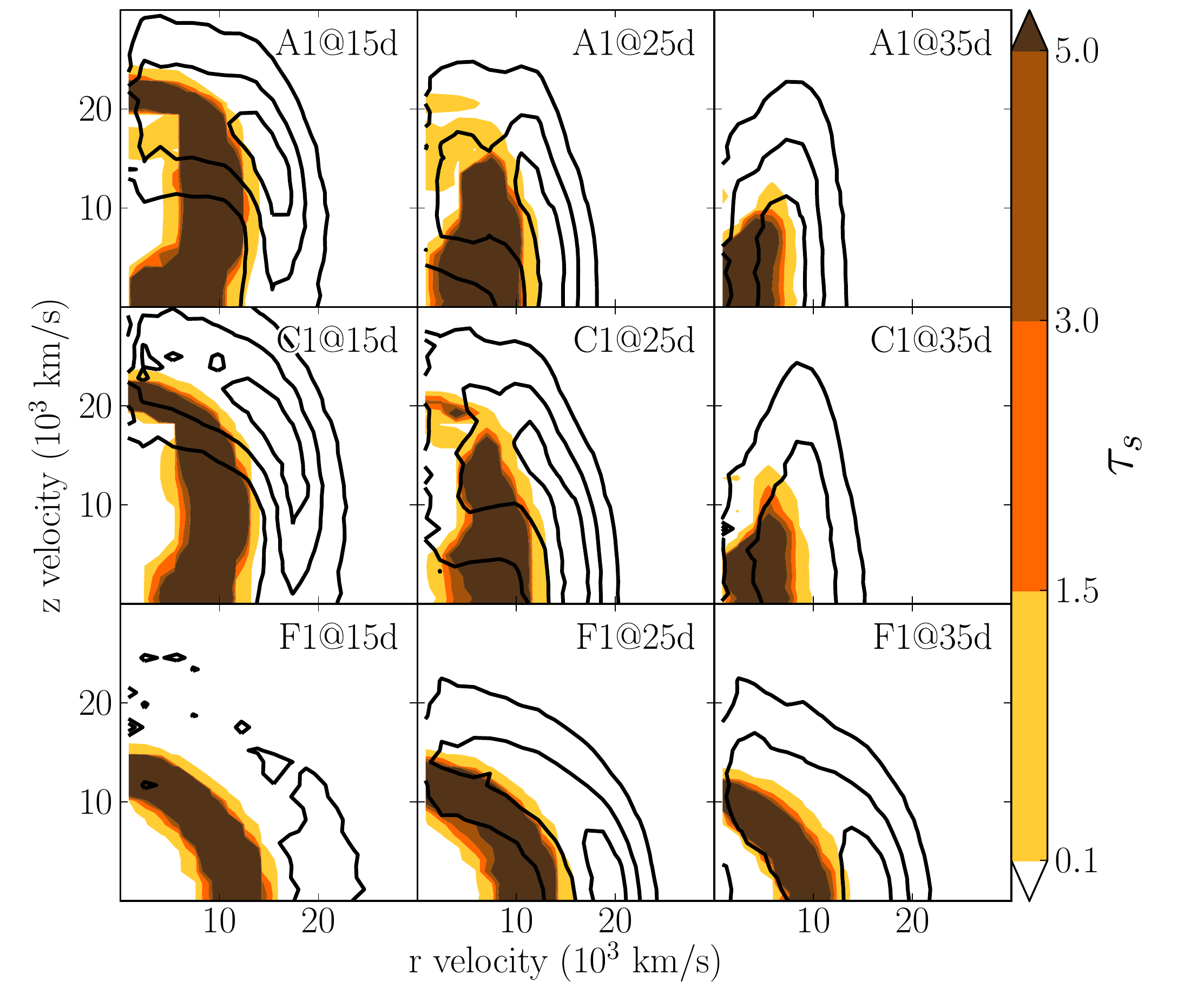}
\caption[short]{Sobolev optical depth (color coded) of the Si~{\sc ii} 6,355\AA \ line
  for models A1, C1 and F1 (from top to bottom) at times 15, 25 and 35
  days after explosion (from left to right). The black contours
  indicate the area of last interaction of escaping R band photons at
  that time (effective photosphere). The contours correspond to 0.2,
  0.5 and 0.8 of the emissivity normalised to peak value.}
\label{fig:SiII}
\end{center}
\end{figure}

As discussed in Section~\ref{light_curve_section}, the SN colors at maximum light
also show a systematic dependence on the observer orientation.
Consequently, our 
simulations predict that the Si velocity and the color 
should be correlated for SNe described by the M02 models.
This predicted correlation is shown in Fig.~\ref{fig:Si_vs_bmv}.
For all our aspherical models, we find that decreasing the inclination
angle causes the peak $B-V$ color to become systematically redder
while the Si~{\sc ii} velocity becomes simultaneously higher. The
range of variation depends on the degree of asphericity. Clearly, the
spherically symmetric models show no variation with observer
orientation while in Model~A1 we find the $B$-$V$ color to differ by
up to $\sim$0.16 magnitudes and the range of Si~{\sc ii}
velocities ($\Delta v_{\text{Si~{\sc ii}}}$) to be
$\sim$6,300 km~s$^{-1}$. Model C1 shows a similar correlation to that
in Model A1, except that the range of variation is smaller since the
degree of asphericity is lower. 

We find that this correlation between line velocity and color is
present in the simulations from $\sim$18 days after explosion 
for as long as the Si line remains in the
spectrum.
However, it changes slightly at different epochs. For
example, in Model~A1, the absolute Si velocities decrease after
maximum light and the amplitude of the color variation with viewing angle becomes
slightly smaller (see Fig.~\ref{fig:BMV_A}). Nevertheless, the
correlation persists and, for 18 days after explosion and all later epochs,
is always in the same sense (redder colors are associated with higher line
velocity). 

\subsection{Na~{\sc i} absorption}

One of the strongest features in our synthetic spectra is 
the Na~{\sc i} 5890\AA \
absorption line (see Fig.~\ref{fig:spectra_features}).
In general, we find this feature to have similar strength and angle dependency
as reported by T07.
The Na in the ejecta is predominantly in the unburned material. Its
abundance is therefore governed by the Na mass-fraction in the 
pre-explosion stellar model and is uniform 
in the outer ejecta (Fig.~\ref{fig:model}). Due to the high optical
depth of this ground state transition, only a small Na mass fraction is
required to make the line strong, if the ionization state is sufficiently low
to favour Na~{\sc i}.  
The angle dependence of the Na line is therefore not a consequence of
the spatial distribution of the Na abundance, but of variations in 
the ionisation
state (see Fig.~\ref{fig:NaINaII}). Particularly for early times, it is
only in the relatively dense equatorial regions that the Na~{\sc i}
population is sufficiently high that the line is very strong (see
Fig.~\ref{fig:NaIwithtime}). However, at later times sufficient recombination
occurs that the line becomes significant for all
orientations. 

\begin{figure}
\begin{center}
\includegraphics[width=\columnwidth]{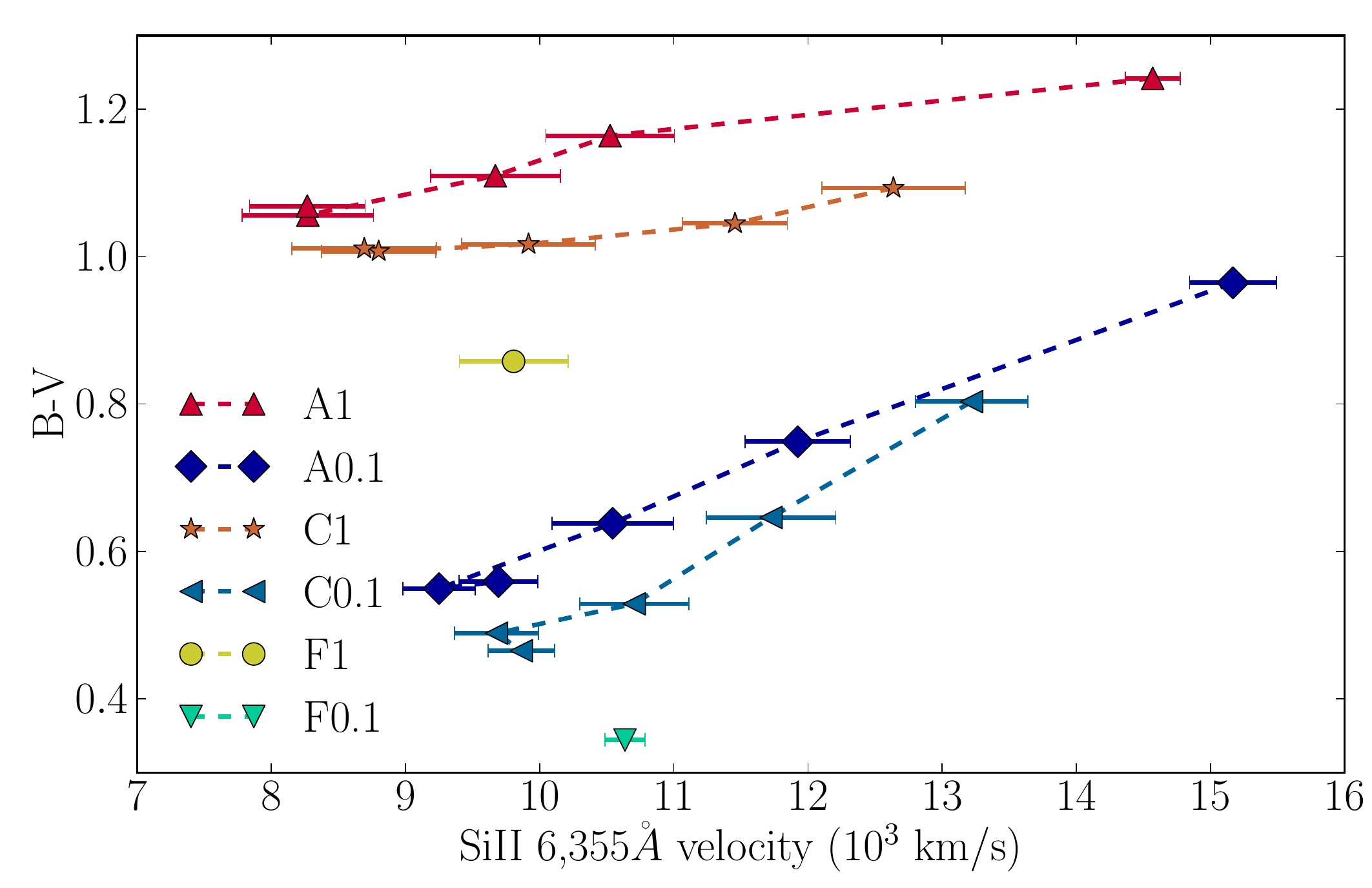}
\caption[short]{$B-V$ color vs.\ Si~{\sc ii} velocity for all the
  models at $V$-Band maximum light (see table \ref{vband} for details). For the aspherical models (A1,A0.1,C1,C0.1) the symbols
  represent data points at 25,45,60,72 and 85 degrees with decreasing Si
  velocity (from right to left). 
The horizontal error bars indicate the uncertainties associated with fitting 
the line velocity from our MC spectra.
The lines connect the data points to
  guide the eye.}
\label{fig:Si_vs_bmv}
\end{center}
\end{figure}

\begin{figure}
\begin{center}
\includegraphics[width=\columnwidth]{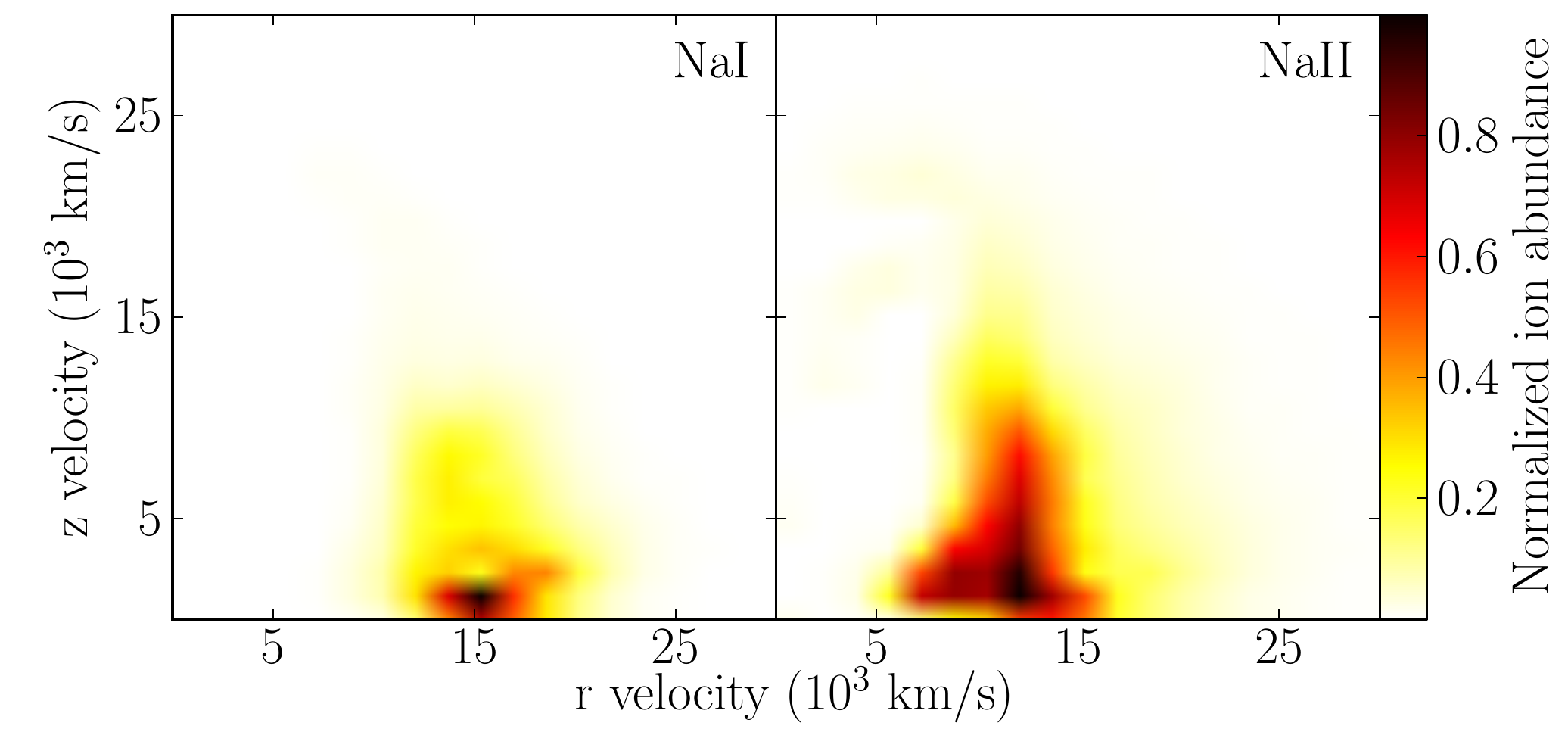}
\caption[short]{Na~{\sc i} (left) and Na~{\sc ii} (right) density distribution for
  model A1 at $V$-Band maximum light (see table \ref{vband} for details). In both panels, the distributions are normalized to a maximum value of 1.0. 
The neutral Na is concentrated along the
  $\theta \sim 90^\circ$ direction explaining the deep absorption observed
from an equatorial orientation.}
\label{fig:NaINaII}
\end{center}
\end{figure}

\begin{figure}
\begin{center}
\includegraphics[width=\columnwidth]{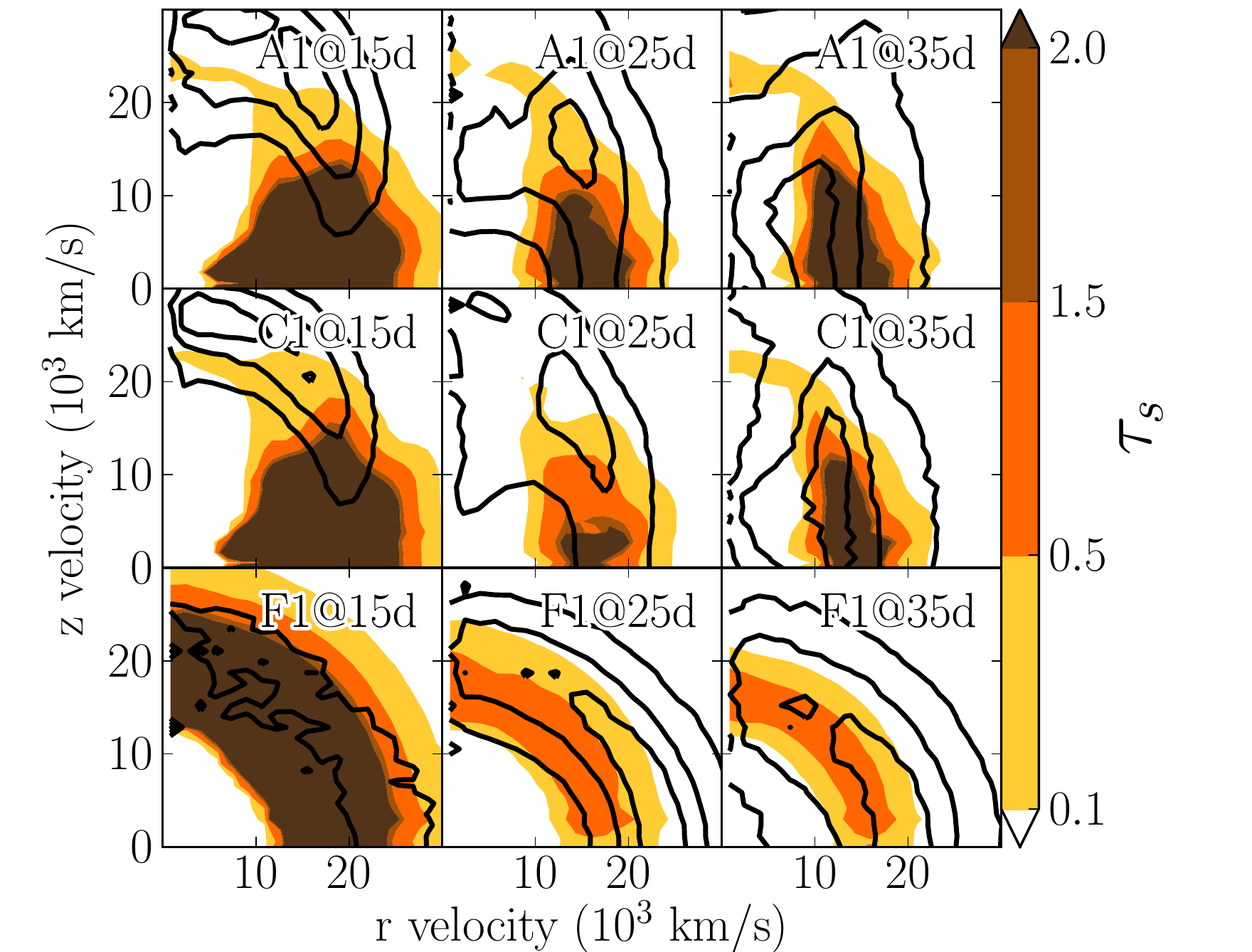}
\caption[short]{Sobolev optical depth (color coded) of the Na~{\sc i} 5,890\AA~ line
  for models A1, C1 and F1 (from top to bottom) at times 15, 25 and 35
  days after explosion (from left to right). The black contours
  indicate the area of last interaction of escaping V band photons at
  that time (efective photosphere). The contours correspond to 0.2,
  0.5 and 0.8 of the emissivity normalised to peak value.}

\label{fig:NaIwithtime}
\end{center}
\end{figure}

As noted by T07, the GRB related SN 1998bw did not show strong Na~{\sc i}
absorption, which is consistent with the models for a 
pole-on viewing angle around maximum light.
However, the 
Na~{\sc i} line remains a challenge for our models 
since the synthetic observables for equatorial orientations are
expected to correspond to cases in which the jet axis is not close to our
line-of-sight. Such off-axis explosions should produce 
a sub-population of SNe~Ic. However, strong Na~{\sc I}
is not typically observed for broad-lined core collapse SNe. 
This may simply be a failing of the particular explosion models and/or 
ionization treatment adopted here.
As discussed above, the formation of this feature is very sensitive to
the degree of ionization in the outer ejecta and it could be
suppressed if the ionization remained slightly higher. Increased
ionisation could be achieved in models with somewhat higher $^{56}$Ni
masses (as would be required to account for the brightness of
e.g. 1998bw \citep{Maeda2006}). 

\subsection{Metallicity Dependence}\label{met}

Studies reveal a possible link between the metallicity
of the SN environment and whether it will produce a GRB
\citep{Sollerman2005,Modjaz2008,Modjaz2006,Levesque2010}. However, conclusive
results are still limited by the small sample of spectroscopically
confirmed  GRB-SNe, only one of which has had sufficiently complete
observations to be modelled in detail (2003dh,
\citealt{Mazzali2003,Matheson2003,Stanek2003,Simon2004,Deng2005}). 
\cite{Sollerman2005} observed three galaxies which were
identified to host a SN with an associated GRB. Their 
results favour low metallicity, sub-luminous
galaxies in a phase of active star formation, which agrees with other 
studies of higher redshift GRBs \citep{LeFloch2003}.
It is therefore important to consider how our synthetic observables
are affected by a reduced progenitor metallicity, to confirm the
robustness of our findings and identify potential observable
signatures of progenitor metallicity in the SN itself.

\begin{figure}
\begin{center}
\includegraphics[width=\columnwidth]{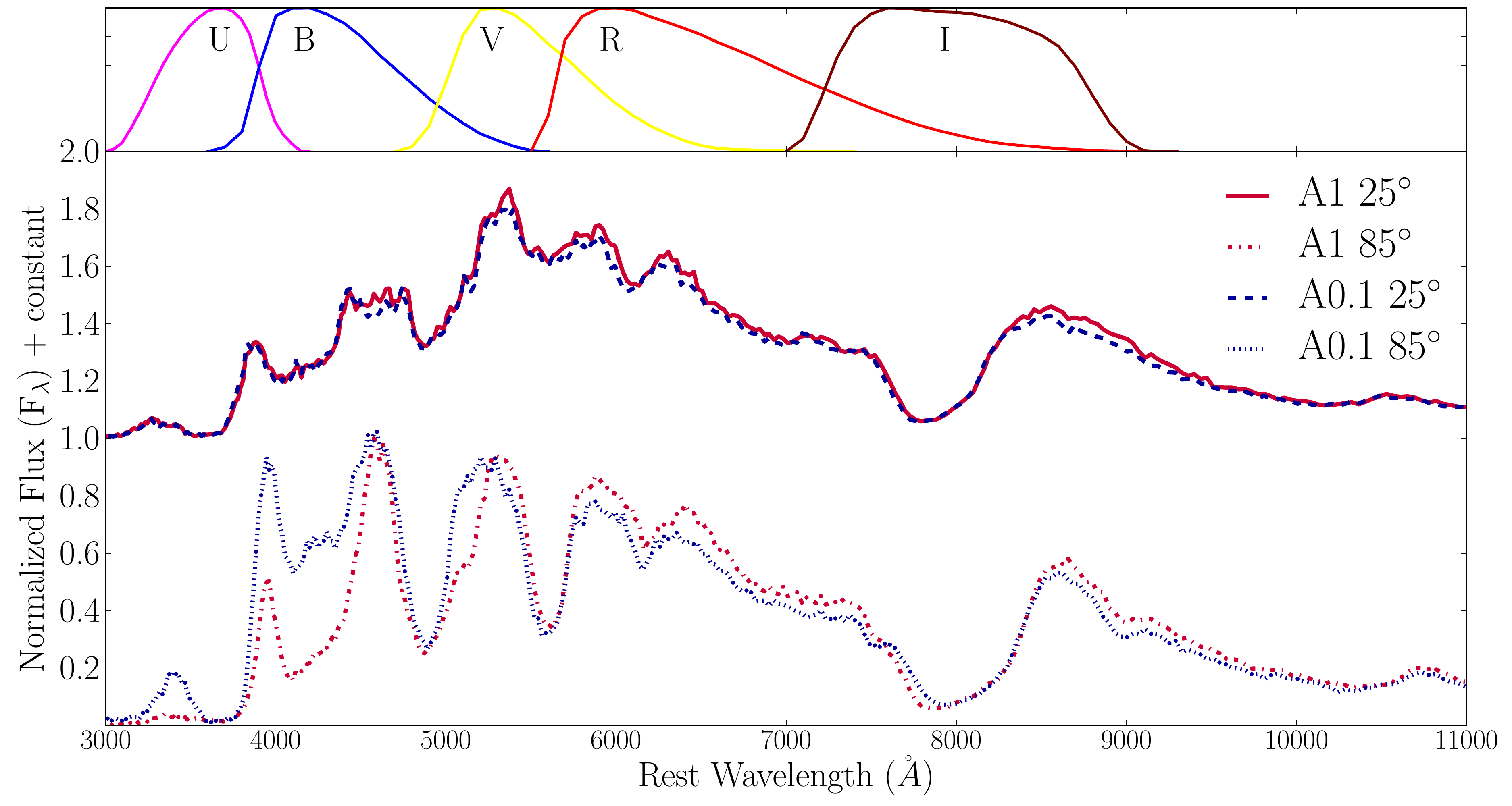}
\caption[short]{Synthetic spectra for models A1 and A0.1 at $V$-band
  maximum light for polar ($\sim25^\circ$) and equatorial
  ($\sim85^\circ$) viewing angles. The top panel shows the filter functions
  used to generate the light curves in Figs.~\ref{fig:A_lc} and \ref{fig:A0.1_lc}.}
\label{fig:A_0p1}
\end{center}
\end{figure}

\begin{figure}
\begin{center}
\includegraphics[width=\columnwidth]{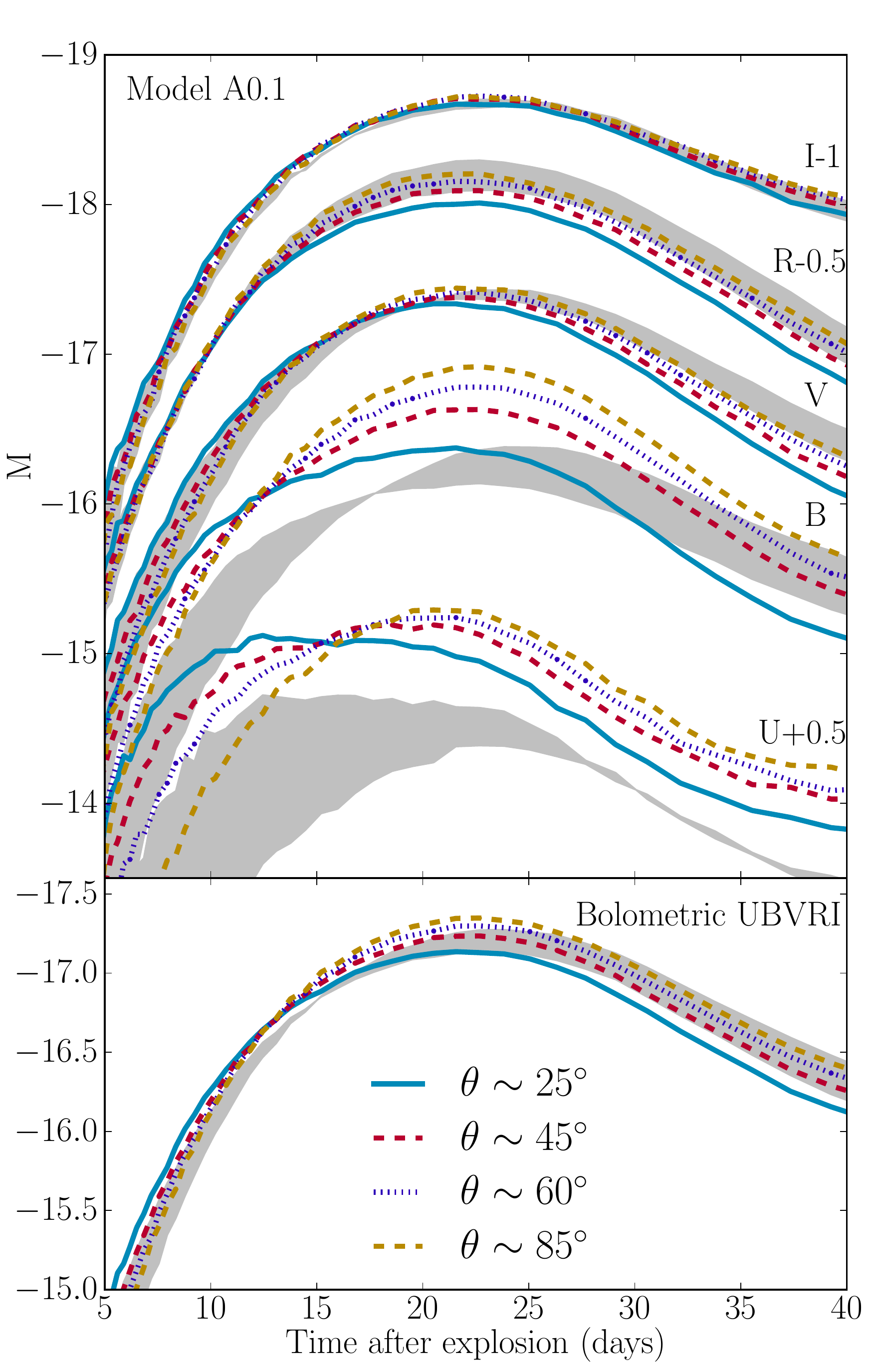}
\caption[short]{Light curves for model A0.1 for the different viewing
  angles. $\theta=0^\circ$ is the jet
direction and $\theta=90^\circ$ is the equator. The shaded gray
area spans the values for all angles for model A1 to allow
comparison. See top panel of
Fig. \ref{fig:spectra_features} for the filter curves.}
\label{fig:A0.1_lc}
\end{center}
\end{figure}

Fig.\ \ref{fig:A_0p1} shows spectra for our sub-solar metallicity
model A0.1 and solar metallicity model A1 for the same epoch as shown
for model A1 in Fig.\ \ref{fig:spectra_features}.
The biggest difference is increased flux in the blue part of the spectrum 
for all orientations, a consequence of less line blocking from primordial
heavy elements in the lower metallicity model. This persists for all
epochs and results in light curves that are systematically brighter up
to $\sim$35 days past explosion and reach peak earlier, in $B$ and $U$ for
model A0.1 (Fig.~\ref{fig:A0.1_lc}) compared to model A1. We draw
similar conclusions when comparing the spherically symmetric
models.

Since the metallicity affects the redder bands much less
significantly, the increased blue flux from the sub-solar metallicity
models influences the colors. The sub-solar models are always bluer
and also show a wider variation of color with orientation. 
This makes the Si~{\sc ii}-color correlation even stronger (and,
consequently, more detectable), spanning a range of $\Delta(B-V)$
$\sim$0.5 around peak (see results for model A1 in
Fig.~\ref{fig:Si_vs_bmv}). The trend in the correlation 
with time is different to that found in model A1 with a plateau around
peak and a beginning of decrease $\sim$35 past explosion.

\section{Discussion/conclusion}\label{conc}

We performed multi-dimensional radiative transfer calculations 
to compute light curves and spectra
for 2D models of SNe driven by bi-polar outflows. 
Compared to previous studies, the main advantage of our work is the
use of self-consistent opacities in a time-dependent  
calculation of the spectral evolution.

As expected for strongly asymmetric explosion models, 
we find that many observable properties depend on the orientation of
the observer.  
During the rise phase, the synthetic light curves are brighter and the
colors are bluer for an observer 
inclination close to the polar axis. However, at peak, the pole-on light curves are slightly fainter and redder than those seen from
an equatorial orientation. This is a consequence of the concentration of $^{56}$Ni along the polar direction in the explosion models.
The time of maximum light also depends slightly on inclination and is
reached a few days earlier for pole-on inclinations, in agreement with the
findings by \cite{Maeda2006}.

The color evolution of our sub-solar metallicity aspherical model A0.1 in the pole-on case shows
similar trends to 1998bw: $B-V$ increases with time, although less
quickly in our model than for 1998bw. 
We note, however, that 1998bw is systematically bluer at
all times by 0.2-0.4 magnitudes.
In our models, we find low metallicity makes the peak $B-V$ color bluer and 
increases the steepness of the $B-V$ temporal
color evolution around peak, 
suggesting that even lower metallicity might be worthy of investigation
for 1998bw.
Comparing our off-axis color evolution to
SN 2009bb, a BL-Ic SN without a GRB, we find similar results (i.e. with
a slower color evolution and redder color in our model than those found by
\cite{Pignata2011} for the SN).

Our asymmetric models predict that an on-axis orientation 
gives rise to a redder $B-V$ color at peak.
This should be statistically testable if a large sample of potential
jet-driven SNe~Ic (including events observed both on- and off-axis)
were identified. This could be most clearly done with a set of 
objects for which the ejecta mass and energy are similar. However,
even for a sample with a range of explosion energies and masses a mean 
trend in the ensemble averages might be detectable.

These orientation-dependent effects on the light curve mean that one would
overestimate the ejected mass and $^{56}$Ni (by on the order of tens per cent)
if applying the widely-used Arnett relation \citep{Arnett1982} to the
equatorially-viewed light curve, and underestimate these parameters
when viewing the SN pole-on.

Spectral features also depend on the viewing angle. As already noted
by T07, the Si~{\sc ii} line velocity at maximum light varies
monotonically (within our MC uncertainties) with orientation, 
being fastest for pole-on
inclinations. We find the Si~{\sc ii} velocities along the poles to be
comparable with those reported for 
SNe associated with GRBs such as 1998bw and 2003lw,
while our off-axis velocities are comparable to BL-Ic SNe such as
1997ef and 2003bg \citep{Corsi2011}. Since our light curve properties also show a simple
trend with viewing angle, this means we 
predict a Si~{\sc ii} velocity -- peak color ($B-V$)
correlation. We found that this correlation is robust:
  it is predicted for solar and sub-solar models during a significant
  range of observable epochs and for a range in asphericities.
Such a correlation
should be statistically testable if a large sample of potential
jet-driven SNe~Ic (including events observed both on- and off-axis)
were identified.

We examined the role of progenitor composition on the light curves and
spectra by considering models appropriate for both solar and one-tenth
solar metallicity. 
We showed that the metallicity strongly influences
the blue part of the spectrum (the $B-V$ color is roughly 0.2 -- 0.8~mag bluer for $Z = 0.1 Z_{\odot}$ compared to $Z =1Z_{\odot}$). 
Therefore, if GRB progenitors typically have low metallicity,
we expect them to have measurably
bluer colors than a typical stripped-envelope core collapse SN with a more metal-rich
progenitor. 
However, the metallicity does not qualitatively affect
the trends of line-velocity and light curve color with orientation
(indeed, we find that our velocity--color relationship is even
stronger for $Z = 0.1 Z_{\odot}$).

There are some shortcomings of our synthetic light curves and spectra
in comparison to observations. Typically, our  
light curves reach peak at $\sim$23 days, which is 7 days later than typical
for SNe associated with GRBs~\citep{Woosley2006} which could imply that
the density distribution in the current models is somewhat 
inappropriate. Also, at most epochs
in our models, we predict strong Na~{\sc i} absorption, 
which is not typical of SNe~Ic. This may be attributed to the
ionisation state being too low in our current models. 
Further study and exploration of different explosion model parameters
is warranted to understand whether these issues can be resolved. 

In this work we used the the well-studied M02 models as a test case
to compare our methods with previous results, and allow us to
understand the limitations and strengths of our approach before
applying them to a wider range of models. We have concentrated on only 
two of the parameters that are relevant to asymmetric Type~Ic
explosions: the degree of asphericity and the composition (i.e. metallicity). However, to fully explore this class of models and make quantitative 
comparisons to observed explosions, we will extend this work to span a range of masses and explosion energies.
Having such a grid of synthetic spectra for models with
different explosion parameters will help to quickly identify parameter ranges for newly observed
GRB-SNe, on which more detailed modelling can be based.

\section*{Acknowledgements}
SR and SAS would like to thank the IPMU for their
hospitality during their visit which resulted in this collaboration. BPS
acknowledges financial support through ARC Laureate Fellowship Grant FL0992131.
SR, SAS and BPS acknowledge The Centre for All-sky Astrophysics is an
Australian Research Council Centre of Excellence, funded by grant
CE110001020. This research was undertaken with the assistance of resources
provided at the NCI National Facility in Canberra, Australia, which is
supported by the Australian Commonwealth Government, through the
National Computational Merit Allocation and the ANU Partner Share Schemes supported by the Australian
Government. This research is supported by the World Premier International Research
Center Initiative (WPI Initiative), MEXT, Japan. K.M. acknowledges
support by Grant-in-aid for Scientific Research (23740141). We would
like to thank the anonymous referee for their constructive comments.

 \bibliographystyle{hapj}

\end{document}